\documentclass[aps,twocolumn,preprintnumbers,nofootinbib,superscriptaddress]{revtex4-2}
\usepackage{eurosym}
\usepackage{amsmath}
\usepackage{graphicx}
\usepackage{amsfonts}
\usepackage{array}
\usepackage{amsthm}
\usepackage{bm}
\usepackage{palatino}
\usepackage{mathpazo}
\usepackage{supertabular}
\usepackage{subfig}
\usepackage[breaklinks]{hyperref}
\usepackage{caption}
\usepackage[font=small,labelfont=bf,justification=raggedright]{caption}
\usepackage{color}
\usepackage{booktabs}
\usepackage{multirow}
\usepackage{amssymb}
\usepackage{graphicx, float}
\usepackage{graphicx, epsfig}
\usepackage{enumerate}
\usepackage{tikz}
\usetikzlibrary{matrix}
\newcommand{\be}{\begin{equation}}
\newcommand{\ee}{\end{equation}}
\newcommand{\ba}{\begin{eqnarray}}
\newcommand{\ea}{\end{eqnarray}}
\newcommand{\bal}{\begin{align}}
\newcommand{\eal}{\end{align}}

\newcommand{\bw}{\begin{widetext}}
\newcommand{\ew}{\end{widetext}}

\usepackage[utf8]{inputenc}
\newcommand{\beq}{\begin{equation}}
\newcommand{\eeq}{\end{equation}}
\newcommand{\bqn}{\begin{eqnarray}}
\newcommand{\eqn}{\end{eqnarray}}

\begin{document}
\title{Joshi--Malafarina--Narayan singularity in weak magnetic field}

\author{Mustapha Azreg-A\"{\i}nou}
\email{azreg@baskent.edu.tr (corresponding author)}
\affiliation{Ba\c{s}kent University, Engineering Faculty, Ba\u{g}l\i ca Campus, 06790-Ankara, Turkey}

\author{Kauntey Acharya}
\email{kaunteyacharya2000@gmail.com}
\affiliation{
International Centre for Space and Cosmology, School of Arts and Sciences,\\
Ahmedabad University, Ahmedabad-380009 (Guj), India.}

\author{Pankaj S. Joshi}
\email{psjcosmos@gmail.com}
\affiliation{
International Centre for Space and Cosmology, School of Arts and Sciences,\\
Ahmedabad University, Ahmedabad-380009 (Guj), India.}

\begin{abstract}
The importance and significance of magnetic fields in the astrophysical scenario is well known. Many domains of astrophysical black hole physics such as polarized shadow image, high energy emitting processes and jet formation are dependent on the behavior of the magnetic fields in the vicinity of the compact objects. In light of this, we determine the master equation and master differential equation that determine the spatial behavior of the magnetic field inside a matter distribution or vacuum region, of general spherically symmetric metric, which is immersed in a test magnetic field. We also investigate here the case of JMN-1 singularity immersed in a uniform weak magnetic field and determine the behavior of magnetic fields by defining electromagnetic four potential vector. We find that the tangential component of the magnetic field is discontinuous at the matching surface of the JMN-1 singularity with the external Schwarzschild metric, resulting in surface currents. We define the covariant expression of surface current density in this scenario. We also analyze the behavior of center-of-mass energy of two oppositely charged particles in the geometry of the magnetized JMN-1 singularity. We briefly discuss the possible scenarios which would possess a discontinuous magnetic field and implications of the same and future possibilities in the realm of astrophysics are indicated.
\end{abstract}

\maketitle

\section{Introduction}

The field of astrophysical black holes is becoming intriguing day by day with existing and upcoming observational data from Event Horizon Telescopes, LIGOs, PTAs and others. It is fascinating to see observatories across the globe capturing electromagnetic observations in different regimes. The domain of black hole imaging and shadows touched the pinnacle with the images of M87{*} and Sgr A{*} released in the last few years. At the same time, phenomena falling within different wavelengths of the electromagnetic spectrum have also been puzzling astronomers and astrophysicists. There have been many attempts to understand the reasons behind such types of observations showing huge amounts of energy output in phenomena like GRBs, FRBs etc. \\

The observation of X-ray source Cygnus X-1 in 1971 got the attention of the physics community, which was widely thought to be a black hole. This embarked interest among researchers to study the possible mechanisms of energy extraction from black holes. The well-known Penrose process proposed to describe the rotational energy extraction from Kerr black hole can be considered as one of the initial works in this field \cite{Penrose:1971uk}. After this, the domain of black holes as sources or emitters of extremely large energy started evolving with many different works. \\

Various mechanisms and processes of energy extraction have been proposed. This also includes attempts to modify the Penrose process by considering different scenarios such as Super-Penrose process, collisional Penrose process etc \cite{Tursunov:2019oiq}. The field of electromagnetic energy extraction started to evolve along with rotational energy extraction. This led to proposals of many sophisticated energy extraction mechanisms such as Blandford-Znajek process \cite{Blandford:1977ds}, Blandford Payne process. Some of these processes and phenomena such as radiation reaction consider the effects of magnetic fields on the dynamics of the charged particles \cite{Tursunov:2018erf}.\\

The topic of inclusion of magnetic fields in astrophysical phenomena dates far back to 1937 when Alfevin proposed the threading of magnetic fields in interstellar space to explain the confinement of cosmic rays. This opened the era of Interstellar Magnetic Fields (ISMFs). Followed by this was published Fermi's work on the origin of cosmic radiation, which considered the concept of moving magnetic fields. While from the observational point of view, Hall and Hiltner  independently discovered linear polarization in the optical light, which was later attributed to Interstellar Magnetic Fields (ISMFs). At about the same time, it was shown that the Synchrotron radiation phenomena can explain the general radio continuum emission from our Galaxy. This process also indicated the existence of magnetic fields in the interstellar medium. The details about this aspect of Interstellar Magnetic Fields (ISMFs) can be found in~\cite{Ferriere_2015} and the role of magnetic fields on accretion disks around rotating black holes and on jets associated with these black holes is detailed in~\cite{Stuchlik2000}.

However, it was not until 1974 that the magnetic fields were considered for black holes when Wald studied electromagnetic fields produced by external sources (e.g., plasma accreting onto the black hole)~\cite{ansatz1}. This idea significantly contributed to the development of many processes that were to be defined in the vicinity of the compact objects. Along with this, the interest in the field of gravitational collapse and visibility of the singularity was also growing. Many studies suggested the existence of singularity without the presence of horizon as a result of gravitational collapse of a matter cloud \cite{Joshi:1993zg, Dwivedi:1994qs, Singh:1994tb}.\\

In recent years, the observational signatures of these singularities is a topic of curiosity. Some of the models of these singularities are considered to be good substitutes of black holes \cite{Saurabh:2023otl,Vagnozzi:2022moj,EventHorizonTelescope:2022xqj}. There have been many  phenomenological works which suggest that, in certain cases, the observational aspects of singularities are very distinct from that of black holes\cite{Bambhaniya:2019pbr,Joshi:2020tlq}. Thus, careful investigation in this direction can provide a way to distinguish singularities from black holes and other compact objects. \\

Other than this, the interest towards the field of interior solutions as opposed to a picture of compact objects with a horizon and singularity is also arising because of multiple reasons. A detailed discussion on this issue is given in the introduction section of \cite{doran2006}. Gravastars or gravitational condensate stars \cite{Mazur2004} , dark energy stars \cite{Chapline2004} etc can be considered to be interior structures. These structures have interior spacetime matched with an exterior Schwarzschild spacetime. The JMN-1 singularity is also an interior solution for which we study the behavior of magnetic fields. Thus, the tangential discontinuity we obtain for JMN-1 singularity can similarly be expected for such interior solutions. \\

In view of this, we investigate the behavior of the magnetic fields in the vicinity of the JMN-1 singularity~\cite{Joshi:2011zm}. JMN-1 singularity forms as an end state of gravitational collapse of a cloud with inhomogenenous matter distribution and tangential pressure. This spacetime is of importance from an astrophysical point of view as it has been considered as one of the best mimickers or substitutes of black holes. To study the behavior of electromagnetic fields in this spacetime, we use a method proposed in ~\cite{ansatz2}. This method is a generalization of Wald's work ~\cite{ansatz1} in which magnetic fields are defined for Schwrazschild black hole and Kerr black hole. In this paper, we also present the generalized formalism to define quantities related to magnetic fields in terms of general metric tensor components for any given general spherically symmetric spacetime.\\

The paper is organized as follows: In Sec.~\ref{JMN-1 Singularity}, we present some details and properties of JMN-1 singularity. In Sec.~\ref{General method}, we present the mathematical formalism for evaluating the electromagnetic four potential vector, electromagnetic field tensor and orthogonal magnetic field components in any given spacetime endowed with spherical symmetry. Following this, we define quantities of magnetic field in JMN-1 spacetime in Sec.~\ref{Magnetic fields in JMN-1}. We further discuss the behavior of the magnetic field in JMN-1 singularity and briefly compare it with magnetic fields in other solutions in Sec.~\ref{Comparison}. In Sec.~\ref{SC}, we define the covariant expression of the surface current at the surface of the interior matter. The energy of center-of-mass is discussed in Sec.~\ref{ECM}. In Sec.~\ref{Discussion}, we summarize our results, discuss notable aspects of this work and its future implications.

\section{JMN-1 Singularity}\label{JMN-1 Singularity}

The foundational study on formation of black hole was done by Oppenheimer, Snyder, and independently by Dutt \cite{Oppenheimer:1939ue}, where it was shown that the gravitational collapse of a spherically symmetric cloud with homogeneous matter distribution leads to formation of a black hole. As a result of gravitational collapse of the matter cloud with zero pressure and constant density, the central spacelike singularity is covered by the trapped surfaces. This suggests existence of an event horizon as the region inside the outermost trapped surface is not causally connected with any other point of the spacetime outside the trapped surface. The well known Cosmic Censorship Conjecture suggests the existence of an event horizon covering the central curvature singularity. However, this model of gravitational collapse can be considered to be a very ideal one because it is assumed that the collapsing star possesses no pressure and the matter distribution is homogeneous. For this, many studies have been carried out in which the gravitational collapse is investigated in more realistic physical scenarios such as  inhomogeneous matter cloud with non zero pressures\cite{Joshi:2011zm}. The consideration of inhomogeneous matter cloud is physically realistic in a sense that the astrophysical stars possess higher density at the centre, decreasing gradually with the radius.\\

These studies suggest that the end state of a star undergoing the process of gravitational collapse can lead to formation of locally or globally visible genuine singularity when the collapsing cloud is considered to be inhomogeneous \cite{Mosani:2022nsp,mosani3}. The gravitational collapse of one of many such scenarios was studied in \cite{Joshi:2011zm} where the cloud is considered to possess tangential pressure components along with the inhomogeneous density. In this case, the anisotropic matter cloud attains the equilibrium configuration suggesting a timelike singularity at the infinite time known as JMN-1 singularity.  The line element of the JMN-1 singularity is,
\begin{equation}
ds^2 = - (1- M_{0}) \left( \frac{r}{r_b}\right)^{\frac{M_{0}}{1 - M_{0}}} dt^2 + \frac{1}{(1- M_{0})}  dr^2 +\, r^2  d\Omega^2,
\label{JMN-1}
\end{equation}
where $d\Omega^2= d\theta^2+\,\sin^2\,{\theta}\,d\varphi^2$. $M_{0}$ and $r_{b}$ are positive constants.The equilibrium energy density and tangential pressure of the anisotropic matter, of which the JMN-1 singularity is made, are given by \cite{Joshi:2011zm}:
\begin{equation*}
\rho_e =\frac{M_0}{r^2},\qquad p_e = \frac{M_0^2}{4(1-M_0)r^2},	
\end{equation*}
which are both positive and diverge in the limit $r\to 0$. The effective sound speed $c$ is such that $c^2=p_e/\rho_e =M_0/[4(1-M_0)]$. Requiring that this effective sound speed be less than the speed of light yields the constraint $0<M_0<4/5$. In the equilibrium configuration, the anisotropic matter is present in the central region of the spacetime, which is represented by the above spacetime \eqref{JMN-1}. In physical sense, the parameter $r_{b}$ is the radius of this matter distribution. The invariant scalar of the above line spacetime given by
\begin{equation}\label{is}
R=\frac{M_0(2-3M_0)}{2(1-M_0)r^2}\,,
\end{equation}
diverges in the limit $r\to 0$ as $1/r^2$. Note that $R$ vanishes for $2/3=M_0<4/5$, but for this same value of $M_0$ the Kretchmann scalar $K$ diverges in the limit $r\to 0$ as $1/r^4$. In fact, $K$ diverges for all $M_0<4/5$ as $r\to 0$:
\begin{equation}\label{ks}
K=\frac{M_0^2(33M_0^2-60M_0+28)}{[2(1-M_0)r^2]^2}\,,
\end{equation}
with $33M_0^2-60M_0+28>0$ for all $M_0$. \\

The above mentioned spacetime represents the central high density region consisting of the anisotropic matter fluid in the equilibrium configuration. Thus, it is necessary to define an exterior spacetime for the region outside the matter cloud such that the structure is asymptotically flat as a whole . Since the radial pressure is zero in JMN-1 the interior spacetime, it is matched with an exterior Schwarzschild spacetime at $r=r_{b}$ radius. The line element of this exterior spacetime can be given by,
\begin{equation}
ds^2=-\left(1-\frac{M_{0}r_{b}}{r}\right)dt^2+\frac{dr^2}{\left(1-\frac{M_{0}r_{b}}{r}\right)}+r^2d\Omega^2,
\label{extsch}
\end{equation}
where, the Schwarzschild mass is given by, $M=M_{0}r_{b}/2$ and $r_{s}=M_{0}r_{b}$ is the Schwarzschild radius.  For the matching of two spacetimes, we mention here junction conditions;\\

(i) It is required for the two spacetimes to match that the induced metrics of internal and external spacetimes are identical on the matching hypersurface. For the case of JMN-1 singularity, it can be seen that this is satisfied for the above metrics \eqref{JMN-1} and \eqref{extsch} at the timelike hypersurface $r=r_{b}$. \\

(ii) Apart from this, it is also required that the extrinsic curvatures $K_{\mu\nu}$ of the two spacetimes also match at the hypersurface. The extrinsic curvature can be defined for any given spacetime by considering the contravariant derivative of the normal vectors on a matching hypersurface. ($K_{ab}=e^\alpha_a e^\beta_b \nabla_\alpha \eta_\beta$, where the normal to the hypersurface is denoted by $\eta_\beta$ while $e^\alpha_a$ and $e^\beta_b$ are tangent vectors on the hypersurface.)\\

Following the above mentioned condition, one can understand that the interior spacetime consisting anisotropic matter and the exterior Schwarzschild spacetime can be matched at $r=r_{b}$. It is also important to note that all the energy conditions are satisfied in case of JMN-1 singularity.\\

The above mentioned JMN-1 singularity is a topic of great importance as it forms an end state of gravitational collapse of matter cloud with physically realistic conditions. Apart from that, there are a number of studies regarding this spacetime which show very exciting results \cite{Shaikh2018lcc,Shaikhhbm}. For this reason, to explore the domain of high energy astrophysics with the possibility of JMN-1 singularity, it is very important to define an appropriate profile of electromagnetic four potential and study the behavior of magnetic fields in the JMN-1 singularity. In the next section, we discuss the derivation of electromagnetic potential $A_\mu$ and electromagnetic field tensor components $F_{\mu\nu}$ for this spacetime.\\

Since our exterior solution is described by Schwarzschild metric immersed in a test magnetic field, from this point of view it is worth mentioning the investigations done in~\cite{PRD,CQG}, which focused on determining the conditions under which a test particle can escape to spatial infinity and on relating the frequencies of the twin high-frequency quasi-periodic oscillations observed in some microquasars to resonant phenomena of the radial and latitudinal oscillations of charged particles, respectively. In our solution the matching surface is located near and outside the horizon: $r_b > r_s=M_0r_b$. However, since the applied magnetic field is assumed to be very weak~\cite{Frolov}, we assume that the ISCO (innermost stable circular orbit) radius is larger than $r_b$: $r_\text{isco}>r_b$, that is, we asuume that the ISCO is located in our exterior region. Under this assumption all conclusions drawn in~\cite{PRD,CQG} remain valid and apply to our exterior metric.

\section{General considerations}\label{General method}
For the purpose of future applications, we first consider the general case of a spherically symmetric metric, of signature ($-,+,+,+$), given by
\begin{equation}\label{m1}
ds^2=g_{tt}(r)dt^2 + g_{rr}(r)dr^2 + g_{\theta\theta}(r)d\Omega^2 .
\end{equation}
Such a metric may describe the interior of a star in its final collapse state, naked singularity, regular or singular black hole, no matter how is the energy-momentum tensor. We assume that the matter distribution has no electric charge distribution with vanishing net electric charge and no internal electric currents. The case with non-vanishing net electric charge could be treated in the same way. For short we call the spherical matter distribution a ``star''.\\

We assume the existence of an external test magnetic field, with magnitude $B$, oriented in the positive $z$ direction at spatial infinity. We aim to determine the magnitude and direction of the magnetic field inside the star. Since there are no electric currents inside the star, this yields  $F^{\mu\nu}{}_{;\nu}=0$ ($\nabla_\nu F^{\mu\nu}=0$) and results in
\begin{equation}\label{m2}
(\sqrt{|g|}F^{\mu\nu}){}_{,\nu}=0.
\end{equation}
In order to solve~\eqref{m2} we need a working ansatz. Our working ansatz is~\cite{ansatz2}
\begin{equation}\label{m3}
A^{\mu}=c_t(r,\theta,B)\xi^{\ \mu}_{t}+\Big[\frac{B}{2}+c_{\varphi}(r,\theta,B)\Big]\xi^{\ \mu}_{\varphi}.
\end{equation}
By this ansatz we admit that the 4-vector potential, $A^{\mu}$, is in the plane spanned by the timelike Killing vector $\xi^{\ \mu}_{t}=(1,0,0,0)$ and spacelike one $\xi^{\ \mu}_{\varphi}=(0,0,0,1)$. This is a generalization of Wald's ansatz~\cite{ansatz1}, where now ($c_t,\,c_{\varphi}$) are no longer constants; rather, they are functions of coordinates and the magnetic field. \\

Using~\eqref{m1} and~\eqref{m3}, we obtain the non-vanishing components of the electromagnetic tensor $F_{\mu\nu}$
\begin{align}\label{m4b}
&F_{t r}=-\partial_r (g_{tt}\,c_t)\,, \qquad F_{t\theta }=-g_{tt}\, \partial_{\theta} c_t \,,\nonumber\\
&F_{r\varphi}=\partial _r \Big[\Big(\frac{B}{2}+c_{\varphi }\Big) g_{\theta\theta}\Big] \sin ^2\theta \,, \\
&F_{\theta\varphi}=\partial _\theta \Big[\Big(\frac{B}{2}+c_{\varphi }\Big) \sin ^2\theta\Big]g_{\theta\theta}  \,.\nonumber
\end{align}
Since there is not electric field inside the star, we must have $F_{t\theta}=-g_{tt}\, \partial_{\theta} c_t=0$ ($\Rightarrow\,c_t$ does not depend on $\theta$) and $F_{t r}=-\partial_r (c_t\,g_{tt})=0$, resulting in
\begin{equation}\label{m5}
c_t(r)=\frac{c}{g_{tt}},
\end{equation}
where $c$ is a constant of integration. Note that the components of the magnetic field, proportional to $F_{\theta\varphi}$ and $F_{r\varphi}$, do not depend on the function $c_t$, so we can take $c=0$ without affecting the subsequent general result.\\

Consider the static observer with four velocity $u^\alpha=e_t=\dfrac{1}{\sqrt{|g_{tt}|}}\,\partial_t$. The other vectors forming the tetrad are $e_r=\dfrac{1}{\sqrt{|g_{rr}|}}\,\partial_r$, $e_\theta=\dfrac{1}{\sqrt{|g_{\theta\theta}|}}\,\partial_\theta$, and $e_\varphi=\dfrac{1}{\sqrt{|g_{\varphi\varphi}|}}\,\partial_\varphi$. This gives ($e_a=h_a{}^\mu\,\partial_\mu$)
\begin{equation}\label{m6}
h_{\hat{t}}{}^t=\frac{1}{\sqrt{|g_{tt}|}}\,,\;h_{\hat{r}}{}^r=\frac{1}{\sqrt{|g_{rr}|}}\,,\;
h_{\hat{\theta}}{}^\theta=\frac{1}{\sqrt{|g_{\theta\theta}|}}\,,\;h_{\hat{\varphi}}{}^\varphi=\frac{1}{\sqrt{|g_{\varphi\varphi}|}}\,,
\end{equation}
and
\begin{equation}\label{m7}
h^{\hat{t}}{}_t=\sqrt{|g_{tt}|}\,,\;h^{\hat{r}}{}_r=\sqrt{|g_{rr}|}\,,\;
h^{\hat{\theta}}{}_\theta=\sqrt{|g_{\theta\theta}|}\,,\;h^{\hat{\varphi}}{}_\varphi=\sqrt{|g_{\varphi\varphi}|}\,.
\end{equation}
The coordinate components of the magnetic field in the frame moving with four velocity $u_\alpha$ read
\begin{equation}\label{m8}
B^\alpha = \frac{1}{2}\,\eta^{\alpha\beta\sigma\mu}F_{\beta\sigma}u_\mu ,
\end{equation}
where $\eta^{\alpha\beta\sigma\mu}$ is the totally antisymmetric tensor $\eta^{\alpha\beta\sigma\mu} = \epsilon^{\alpha\beta\sigma\mu}/\sqrt{|g|}$, with $\epsilon^{\alpha\beta\sigma\mu}$ being the totally antisymmetric symbol ($\epsilon^{tr\theta\varphi}=+1$). Then, the orthogonal components are given by
\begin{equation}\label{m9}
(B_a=) B^a = h^a{}_\mu\,B^\mu .
\end{equation}
Using~\eqref{m8} and the expression of $F_{r\varphi}$~\eqref{m4b}, we obtain
\begin{equation}\label{m10}
B^\theta = 2\,\frac{1}{2}\,\frac{\epsilon^{\theta r\varphi t}}{\sqrt{|g|}} F_{r\varphi}u_t =-	\frac{\partial _r \Big[\Big(\frac{B}{2}+c_{\varphi }\Big) g_{\theta\theta}\Big] \sin\theta}{g_{\theta\theta}\sqrt{|g_{rr}|}},
\end{equation}
and
\begin{equation}\label{m11}
B^{\hat{\theta}}=h^{\hat{\theta}}{}_\theta B^\theta =  - \partial _r (g_{\theta\theta}C_\varphi )\frac{\sin\theta}{\sqrt{|g_{\theta\theta}g_{rr}|}}\,,
\end{equation}
where $C_\varphi =c_\varphi +(B/2)$. In the same manner, we obtain
\begin{equation}\label{m12}
B^r = \frac{(B+2
	c_{\varphi}) \cos\theta +\sin\theta\, \partial _{\theta} c_{\varphi}}{\sqrt{|g_{rr}|}},
\end{equation}
and
\begin{equation}\label{m13}
B^{\hat{r}}=h^{\hat{r}}{}_r B^r = 2C_{\varphi} \cos\theta +\sin\theta\, \partial _{\theta} C_{\varphi}=\frac{\partial_\theta(C_{\varphi} \sin^2\theta)}{\sin\theta}\,.
\end{equation}

Note that for the Schwarzschild metric $c_\varphi \equiv 0$ ($C_\varphi =B/2$)~\cite{ansatz1}, the right-hand sides of~\eqref{m11} and~\eqref{m13} reduce to the usual Newtonian values for $B^{\hat{r}}$ and $B^{\hat{\theta}}$ in the limit $r\to\infty$.\\

There remains to determine the function $c_\varphi(r,\theta)$ by solving the only remaining equation $(\sqrt{|g|}F^{\varphi\nu}){}_{,\nu}=0$, as the other three equations $(\sqrt{|g|}F^{\mu\nu}){}_{,\nu}=0$ ($\mu: t,\,r,\,\theta$) are already satisfied even with $c\neq 0$ in~\eqref{m5}. Equation $(\sqrt{|g|}F^{\varphi\nu}){}_{,\nu}=0$ reduces to the following partial differential equation where $C_\varphi =c_\varphi +(B/2)$.
\begin{multline}\label{L1}
\partial_r\Big(\sqrt{-\frac{g_{tt}}{g_{rr}}}\,\partial_r (g_{\theta\theta}C_\varphi)\Big)\sin\theta = \\
-\partial_\theta\Big(\frac{1}{\sin\theta}\partial_\theta (\sin^2 \theta C_\varphi)\Big)\sqrt{-g_{tt}g_{rr}}\,.
\end{multline}
The general solution may be expanded in terms of Legendre polynomials, $P_n(\cos\theta)$, as
\begin{equation}\label{L2}
C_\varphi(r,\theta) =- \sum_{n=1}f_n(r)\, \frac{\partial_\theta P_n(\cos\theta)}{\sin\theta}\,,
\end{equation}
yielding the differential equation for the functions $f_n(r)$
\begin{equation}\label{L3}
\partial_r\Big(\sqrt{-\frac{g_{tt}}{g_{rr}}}\,\partial_r (g_{\theta\theta}f_n)\Big) = n(n+1)\sqrt{-g_{tt}g_{rr}}\,f_n\,.	
\end{equation}
In deriving~\eqref{L3} we have used the recurrent relations between the Legendre polynomials~\cite{handbook}: {\small $$\frac{n}{n-1}P_n(\cos\theta)-\frac{2n-1}{n-1}\cos\theta P_{n-1}(\cos\theta)+P_{n-2}(\cos\theta)=0.$$}

\subsection{Application: Magnetic fields in JMN-1 Singularity}\label{Magnetic fields in JMN-1}
The line element of the interior spacetime is defined by Eq.~\eqref{JMN-1}. We emphasize again that the interior of the star is not considered to be electrically charged at any given point during the gravitational collapse or at the equilibrium state. For this reason, we can consider that there do not exist electric currents in the interior of the collapsing cloud. With that said, we straightforwardly apply the results of the previous section.\\

We seek an interior solution for the magnetic field such that the radial (normal to the surface $r=r_b$) component of the magnetic field can be matched to the corresponding expression for the Schwarzschild metric, where $B^{\hat{r}}=B\cos\theta$ ($r>r_b$), without having surface magnetic charges. According to Eq.~\eqref{m13}, such matching of the internal and external radial components, $B^{\hat{r}}$, at $r=r_b$ for all $\theta$ is possible only if $C_\varphi$ were independent of $\theta$: $C_\varphi(r,\theta)\equiv C_\varphi(r)$. Since $P_1(\cos\theta)=\cos\theta$, this means we have to keep only the term $n=1$ in Eq.~\eqref{L2}. In this case, Eq.~\eqref{L3} reduces to a homogeneous Cauchy-Euler differential equation [$f_1(r)=C_\varphi(r)$]:
\begin{equation}\label{e2}
2 (1-M_0) r^2 C_{\varphi }'' + (8-7 M_0) r C_{\varphi }' - 2 M_0 C_{\varphi }  = 0\,,
\end{equation}
Where `prime' denotes derivative with respect to $r$. The general solution is given by
\begin{equation}\label{Fuv,phi solution}
C_{\varphi }(r)=c_1 r^{\alpha} +c_2 r^{\beta}\,,
\end{equation}
where ($c_1,\,c_2$) are constants of integration and
\begin{align}
\label{alpha}&\alpha = \frac{5 M_0-6+\sqrt{36-44 M_0+9 M_0^2}}{4 (1-M_0)},\\
&\beta = \frac{5 M_0-6-\sqrt{36-44 M_0+9 M_0^2}}{4 (1-M_0)},
\end{align}
with $36-44 M_0+9 M_0^2>0$ for $M_0<4/5$.\\

Since the value of the metric parameter $M_0$ can be in the range $0<M_0<4/5$, the range of newly defined parameter $\alpha$ is between $0<\alpha<0.70156$ and the range of newly defined parameter $\beta$ is between $-5.70156<\beta<-3$.\\

From this, to obtain appropriate expressions of magnetic field components, we need to define the integration constants $c_1$ and $c_2$ in~\eqref{Fuv,phi solution}. Here, it is also important to remember that the expression~\eqref{Fuv,phi solution} is only for the interior spacetime defined by~\eqref{JMN-1} up to $r=r_b$. The exterior geometry is represented by Schwarzschild spacetime, for which, there exists well defined expressions of electromagnetic field tensor and electromagnetic four potential in the literature~\cite{ansatz1}. \\

Since $\beta <0$, we choose to take $c_2=0$ to have finite values of the orthogonal components of the magnetic field~\eqref{m11} and~\eqref{m13} at $r=0$. To fix the value of $c_1$, the normal component $B_{\hat{r}}$~\eqref{m13} must be continuous (by Gauss theorem) across the surface at $r=r_b$; for otherwise, we would have surface magnetic charges. Thus, we must have (recall $\partial_\theta c_\varphi =0$)
\begin{equation}\label{e5}
C_{\varphi}(r=r_b)=\frac{B}{2},
\end{equation}
yielding
\begin{equation}\label{e6}
c_1 r_b^{\alpha} =\frac{B}{2}.
\end{equation}
(Note that in the external Schwarzschild region we have $c_{\varphi }\equiv 0$ ($C_{\varphi}=B/2$)~\cite{ansatz1}). Finally, we obtain
\begin{equation}\label{e7}
C_{\varphi }(r)=
\begin{cases}
	\dfrac{B}{2} \Big(\dfrac{r}{r_b}\Big)^{\alpha}  & (r\leq r_b),\vspace{3mm}\\
	\dfrac{B}{2} & (r>r_b),
\end{cases}
\end{equation}
and Eq.~\eqref{m3} reads
\begin{equation}\label{final Au}
A^{\mu}(r) =
\begin{cases}
	\dfrac{B}{2}\Big(\dfrac{r}{r_b}\Big)^{\alpha}\xi^{\ \mu}_{\varphi} & (r\leq r_b),\vspace{3mm}\\
	\dfrac{B}{2}\,\xi^{\ \mu}_{\varphi} & (r>r_b).
\end{cases}
\end{equation}

Equations~\eqref{m11} and \eqref{m13} read
\begin{equation}\label{e9}
B_{\hat{\theta}} =
\begin{cases}
	-\dfrac{2+\alpha}{2}\,\sqrt{1-M_0}\,B\Big(\dfrac{r}{r_b}\Big)^{\alpha}\sin\theta & (r\leq r_b),\vspace{3mm}\\
	-\sqrt{1-\dfrac{M_0r_b}{r}}\,B \sin\theta & (r>r_b),
\end{cases}
\end{equation}
\begin{equation}\label{e10}
B_{\hat{r}} =
\begin{cases}
	B\Big(\dfrac{r}{r_b}\Big)^{\alpha}\cos\theta & (r\leq r_b),\vspace{3mm}\\
	B \cos\theta & (r>r_b).
\end{cases}
\end{equation}

Note that for the Schwarzschild case ($r>r_b$), we find the usual Newtonian values in the limit $r\to\infty$ ($B$ is oriented in the $+z$ direction).\\

\begin{figure*}
\includegraphics[width=.33\textwidth]{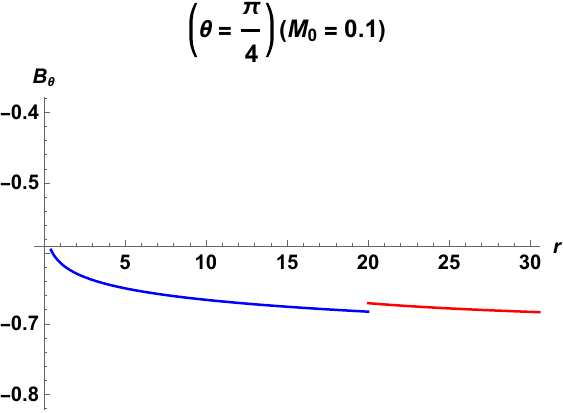}
\includegraphics[width=.33\textwidth]{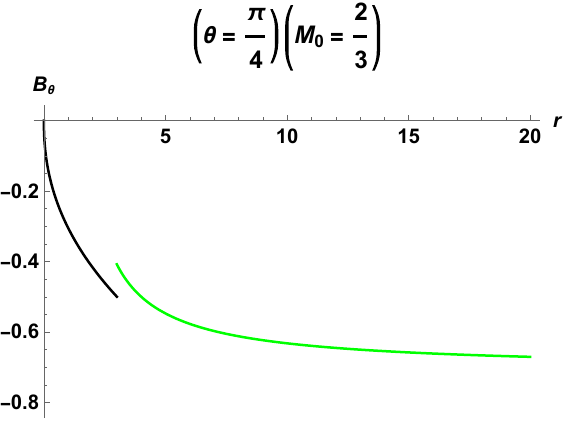}
\includegraphics[width=.33\textwidth]{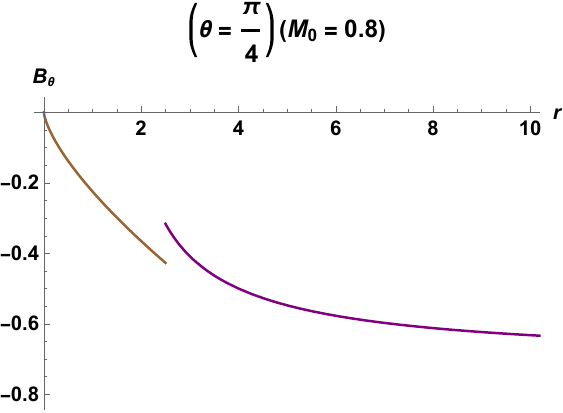}    \includegraphics[width=.33\textwidth]{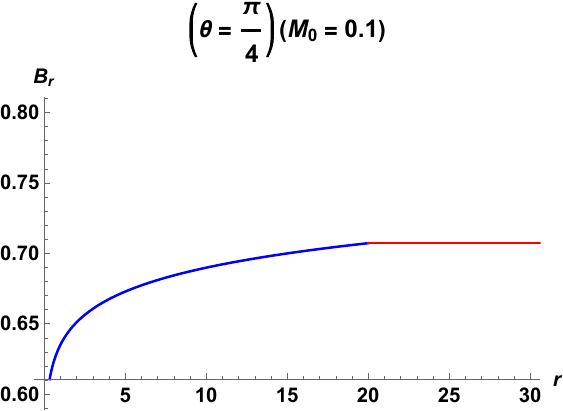}    \includegraphics[width=.33\textwidth]{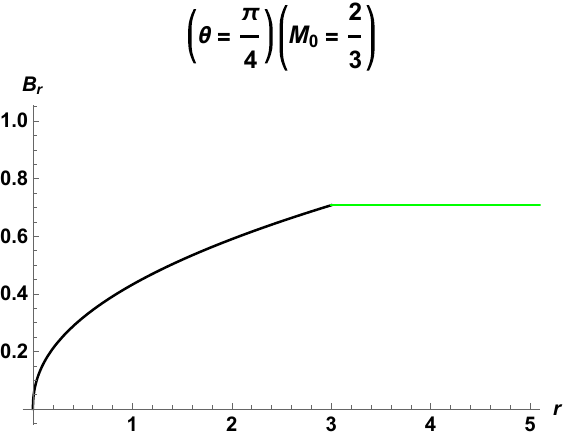}
\includegraphics[width=.33\textwidth]{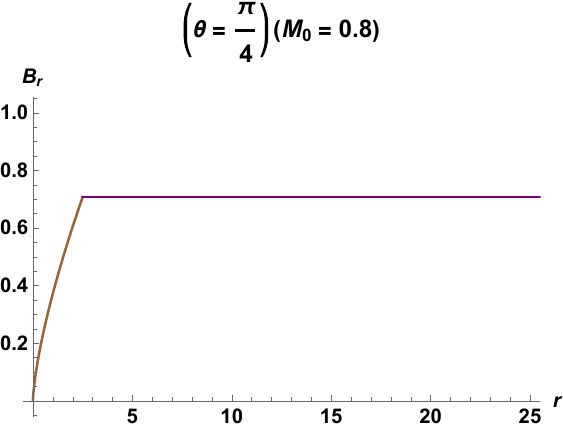}
\caption{Radial and Tangential Components of Magnetic field in JMN-1 Singularity at $\theta=\frac{\pi}{3}$ for $M_0=0.1,2/3,0.8$. Both orthogonal components of the magnetic field for the interior and exterior Schwarzschid spacetime represented by blue and red line for  $M_0=0.1$, black and green line for  $M_0=2/3$, and brown and purple line for  $M_0=0.8$ respectively. The other parameters are $M=1$, $r_b=2/M_0$, and $B\cos(\pi/4)=0.7$. }\label{fig.2}
\end{figure*}

From Eq.~\eqref{e9}, we see that the tangential component, $B_{\hat{\theta}}$, is discontinuous across the boundary surface $r=r_b$. This type of discontinuities are known in magnetostatics and imply the existence of surface currents in the $\varphi$ direction. The value of the surface current depends on the material content of the fluid making up the matter distribution. In the next section \ref{SC}, we define the covariant expression of the surface current. The behavior of these orthogonal magnetic field components with radial distance coordinate is shown in Fig.~\ref{fig.2} for different values of the metric parameter $M_0$ at $\theta=\frac{\pi}{3}$ for the interior spacetime consisting matter and exterior Schwrazschild spacetime.\\

We can write the non-vanishing components of the electromagnetic field tensor $F_{\mu\nu}$~\eqref{m4b} for the interior spacetime $(r\leq r_b)$ as
\begin{equation}\label{Frphi final}
F_{r\varphi}=\frac{2 + \alpha}{2} B r \left(\frac{r}{r_b}\right)^\alpha  \sin^2{\theta},
\end{equation}
\begin{equation}\label{Ftheta phi final}
F_{\theta \varphi}= B r^2 \left(\frac{r}{r_b}\right)^\alpha\sin{\theta} \cos{\theta}.
\end{equation}
%

In the above expressions, the value of $\alpha$ is in the range $0<\alpha<0.701562$ as the metric parameter $M_0$ can have any value from $0<M_0<\frac{4}{5}$.\\

From the above expressions of the electromagnetic four potential~\eqref{final Au}, electromagnetic field tensor components~\eqref{Frphi final} and~\eqref{Ftheta phi final}, and components of the magnetic field~\eqref{e9} and~\eqref{e10}, one can observe that these quantities in the JMN-1 spacetime depend on the newly defined parameter $\alpha$ ($0<\alpha<0.701562$) or the spacetime metric parameter $M_0$ ($0<M_0<4/5$). 


\section{Comparison of behavior of magnetic fields in different geometries }\label{Comparison}
In the previous section, we defined electromagnetic four potential, electromagnetic field tensor and orthogonal components of the magnetic field in JMN-1 spacetime. We also showed that, for the physically realistic behavior of magnetic fields in JMN-1 singularity, the tangential component of the magnetic field is discontinuous at the boundary radius or the matching radius. This type of behavior of the magnetic field has not been observed in the context of compact objects in previous studies. In~\cite{ansatz2}, magnetic field profiles for following different non-rotating geometries were studied: (1) Schwarzschild black hole spacetime (2) Regular Schwarzschild MOG spacetime (3) Reissner-Nordstr\"om spacetime (4) Reissner-Nordstr\"om MOG spacetime (5) Phantom Reissner-Nordstr\"om spacetime. Apart from this, in ~\cite{Hayward,Gammametric,ErnstSch}, magnetic fields are defined for Hayward regular black hole, $\gamma$-metric and Ernst Schwarzschild spacetime, which possess continuous matter fluids. For these spacetimes with continuous matter fluids, as there are no matching surfaces, the orthogonal components of the magnetic field are continuous (in regions of space where no physical singularities occur). Discontinuities in the magnetic field are mostly observed on the surfaces of stars~\cite{sun}.\\

Thus, the discontinuity in the tangential magnetic field component is a distinguishable feature of JMN-1 singularity. The primary reason for such discontinuity is associated with the structure of JMN-1 singularity. As described previously, JMN-1 singularity has a central region consisting of anisotropic matter fluid up to matching radius or boundary radius $r_b$ and an exterior region with vacuum. According to magnetostatics, magnetic fields can be discontinuous at the boundary separating two different media having differences in their properties such as permeability. From this, one can understand that, in JMN-1 singularity, the medium of the interior spacetime ($r<r_b$) is anisotropic matter fluid. While the medium in the exterior spacetime ($r>r_b$) is vacuum. The implication of this discontinuity may emerge when phenomena related to magnetic fields in JMN-1 singularity will be investigated. In this scenario, one can expect notable differences in certain physical properties such as intensity of the radiation at the boundary radius. An instance of that is flaring phenomena on the surface of the Sun, which are strongly believed to be due to existing discontinuities in the surface magnetic field~\cite{sun}.\\

Apart from this, we also note that this kind of discontinuous behavior of the tangential component of magnetic field can be present in any structure in which two different spacetimes are matched or there exists a boundary in the structure which separates two different media. Thus, this property can be useful to distinguish compact objects having continuous fields from those which do not possess such continuity by carefully analyzing the observational data of phenomena which associate effects of magnetic fields.

\section{Covariant expression of Surface Current} \label{SC}

In the previous section, we have mentioned that the discontinuity in the tangential component of the magnetic field suggests the existence of surface current at the boundary separating two different media. In this section, we define covariant expression of the surface current at the boundary of the JMN-1 interior cloud consisting anisotropic matter fluid.\\

For this scenario, the surface current density is denoted by $K$. The boundary surface $S$ separates the vacuum medium having permeability $\mu_{\text{Sch}}=\mu_0$ from the interior medium of permeability $\mu_{\text{jmn1}}$. The covariant expression of $K$ can be obtained from the projection operator,

\begin{equation}\label{p1}
H^{\alpha\beta}=g^{\alpha\beta}-n^\alpha n^\beta\,,\qquad H^\alpha_\beta = \delta^\alpha_\beta - n^\alpha n_\beta	\,,
\end{equation}
where $n^\alpha$ is the unit normal to the surface $S$. In classical electrodynamics, the expression of the surface current is defined as $K=H_1-H_2$ while $H=\frac{B}{\mu}$ . From \eqref{p1}, The component of the magnetic field parallel to the surface $S$: $\Psi(r,\theta,\varphi)=0$ is given by
\begin{equation}\label{p2}
B^{\parallel\,\alpha}=H^\alpha_\beta B^{\beta}=B^{\alpha}-(B^\beta n_\beta)n^\alpha \,.
\end{equation}
From this, the covariant expression of surface current density can be written as,
\begin{equation}\label{p3}
K_\mu = \eta_{\rho\sigma\nu\mu}n^\sigma \Big(\frac{B_{\text{1}}}{\mu_{\text{1}}} - \frac{B_{\text{2}}}{\mu_{\text{2}}}\Big)^{\parallel\,\nu} u^\rho\,,
\end{equation}
where $u^\rho$ is the four velocity and $\eta_{\rho\sigma\nu\mu}$ is the antisymmetric tensor $\eta_{\alpha\beta\sigma\mu} = \sqrt{|g|}\,\epsilon_{\alpha\beta\sigma\mu}$, with $\epsilon_{\alpha\beta\sigma\mu}$ being the totally antisymmetric symbol ($\epsilon_{tr\theta\varphi}=+1$). In the above expression of covariant surface current, the unit normal to the surface $S$: $\Psi(r,\theta,\varphi)=0$ is defined by
\begin{equation}\label{p4}
n_\sigma=\frac{\partial_\sigma\Psi}{\sqrt{g^{\mu\nu}\partial_\mu\Psi\partial_\nu\Psi}}\, ,
\end{equation}
\begin{equation}\label{p5}
\hspace{-3mm}\Big(\frac{B_{\text{1}}}{\mu_{\text{1}}} - \frac{B_{\text{2}}}{\mu_{\text{2}}}\Big)^{\parallel\,\nu}=\Big(\frac{B_{\text{1}}}{\mu_{\text{1}}} - \frac{B_{\text{2}}}{\mu_{\text{2}}}\Big)^{\nu}-\Big[\Big(\frac{B_{\text{1}}}{\mu_{\text{1}}} - \frac{B_{\text{2}}}{\mu_{\text{2}}}\Big)^{\beta}n_\beta\Big]n^\nu .
\end{equation}
If the surface $S$ has the equation $\Psi(r)=0$, which is the case in astrophysical applications, then Eq.~\eqref{p3} reduces to
\begin{equation}\label{k2}
K_\mu =\eta_{\rho\sigma\nu\mu}n^\sigma \Big(\frac{B_{\text{1}}}{\mu_{\text{1}}} - \frac{B_{\text{2}}}{\mu_{\text{2}}}\Big)^{\nu}u^\rho\,.
\end{equation}

The above defined expression of covariant surface current Eq.~\eqref{p3} is the cross product of the unit normal to $S$ and the tangential component of $B^\nu$. It
applies to any surface $S$ separating two different media having permeabilities $\mu_{1}$ and $\mu_{2}$. For the specific case of JMN-1 singularity, the covariant expression of surface current can be defined as
\begin{multline}\label{p6}
K_\mu = \eta_{\rho\sigma\nu\mu}n^\sigma \Big(\frac{B_{\text{sch}}}{\mu_{\text{Sch}}} - \frac{B_{\text{jmn1}}}{\mu_{\text{jmn1}}}\Big)^{\parallel\,\nu} u^\rho =\\ \eta_{\rho\sigma\nu\mu}n^\sigma \Big(\frac{B_{\text{sch}}}{\mu_{\text{Sch}}} - \frac{B_{\text{jmn1}}}{\mu_{\text{jmn1}}}\Big)^{\nu} u^\rho\,.
\end{multline}

There are several papers in which the discontinuous behavior of the magnetic fields are observed in astrophysical bodies such as neutron stars. In Ref.~\cite{Gralla2014, Gurevich, deSouza2018}, the authors defined surface currents on the surface of neutron stars as a result of discontinuity in the magnetic fields. In a similar way, for any massive star undergoing catastrophic gravitational collapse, this can lead to end states with regular structure which can be termed as interior solutions. In these types of structures, the tangential component of the magnetic field would be discontinuous yielding surface currents.

\begin{figure*}
\includegraphics[width=.44\textwidth]{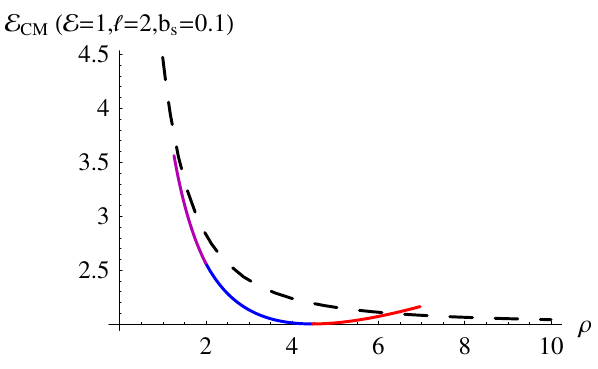}
\includegraphics[width=.44\textwidth]{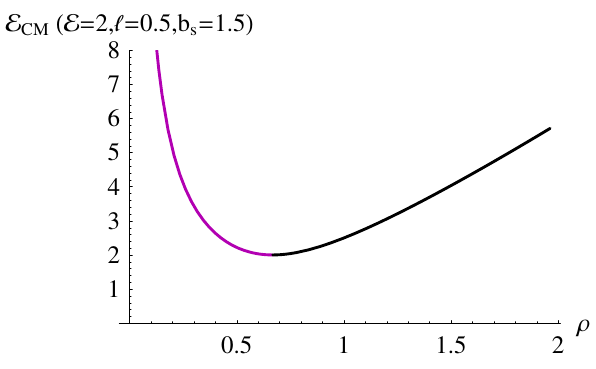}
\caption{Center-of-mass energy versus the dimensionless radial coordinate for $M_0=0.5$. For a Schwarzschild black hole immersed in a magnetic field, a blue (resp. red) plot corresponds to repulsive (resp. attractive) motion of the two particles. For the JMN-1 naked singularity, a purple (resp. black) plot corresponds to repulsive (resp. attractive) motion of the two particles. The motion of the particles is confined as the right-hand side of~\eqref{F1} and of~\eqref{F3} has to be positive (see text for more details). A dashed plot, which has the same shape for all values of the parameters, corresponds to a JMN-1 singularity (internal metric) and a Schwarzschild black hole (external metric) with no external magnetic field ($b_s=B=0$); this plot diverges as $\rho\to 0$ and extends from $\rho=0$ to $\infty$.}
\label{figCM1}
\end{figure*}
\begin{figure*}
\includegraphics[width=.44\textwidth]{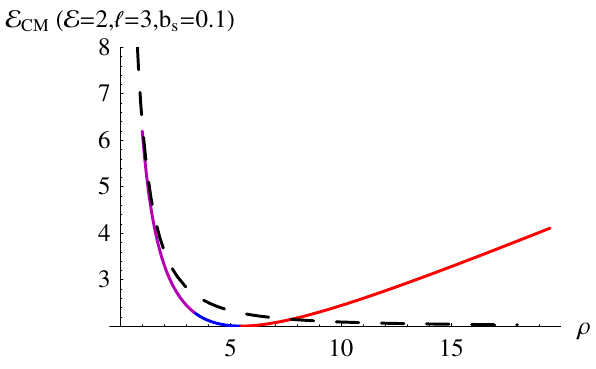}
\includegraphics[width=.44\textwidth]{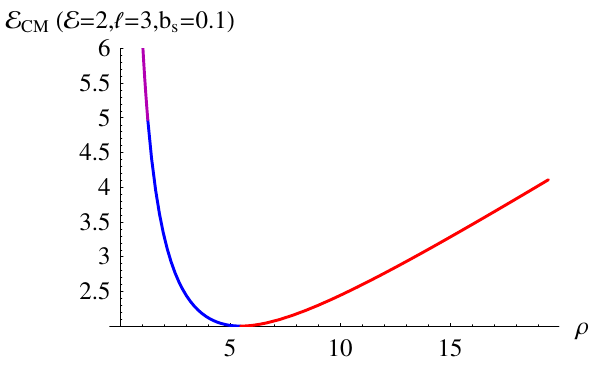}
\caption{Center-of-mass energy versus the dimensionless radial coordinate for $M_0=0.3$ (left panel) and $M_0=0.8$ (right panel). For a Schwarzschild black hole immersed in a magnetic field, a blue (resp. red) plot corresponds to repulsive (resp. attractive) motion of the two particles. For the JMN-1 naked singularity, a purple (resp. black) plot corresponds to repulsive (resp. attractive) motion of the two particles. The motion of the particles is confined as the right-hand side of~\eqref{F1} and of~\eqref{F3} has to be positive (see text for more details). A dashed plot, which has the same shape for all values of the parameters, corresponds to a JMN-1 singularity (internal metric) and a Schwarzschild black hole (external metric) with no external magnetic field ($b_s=B=0$); this plot diverges as $\rho\to 0$ and extends from $\rho=0$ to $\infty$.}
\label{figCM2}
\end{figure*}

\section{Application: Center-of-mass energy}\label{ECM}

First we consider the motion of a charged particle with mass $m$, charge $q$, energy $E$, and angular momentum $L$ in the $\theta=\pi/2$ plane. The case of a Schwarzschild black hole immersed in an external magnetic field has been treated in~\cite{Frolov} where the relevant equations of motion, expressed in terms of the dimensionless coordinates and parameters
\begin{multline}\label{F0}
\rho =\frac{r}{r_s},\quad \sigma =\frac{\tau}{r_s},\quad \ell =\frac{L}{mr_s},\\
\mathcal{E} =\frac{E}{m},\quad b_s=\frac{qBr_s}{2m},\quad r_s=2M,
\end{multline}
where $\tau$ is the proper time, take the following form 
\begin{align}
\label{F1}&\Big(\frac{d\rho}{d\sigma}\Big)^2=\mathcal{E}^2-\Big(1-\frac{1}{\rho}\Big)(1+\beta_{\text{Sch}}^2)\,,\\
\label{F2}&\rho\,\frac{d\varphi}{d\sigma}=\beta_{\text{Sch}}\,,\qquad \beta_{\text{Sch}}=\frac{\ell}{\rho}-b_s\rho\,.
\end{align}

These equations generalize to the case of the JMN-1 naked singularity as
\begin{align}
\label{F3}&\Big(\frac{d\rho}{d\sigma}\Big)^2=\frac{\mathcal{E}^2}{(-g_{tt}g_{rr})}-\frac{1}{g_{rr}}(1+\beta_{\text{jmn1}}^2)\,,\\
\label{F4}&\rho\,\frac{d\varphi}{d\sigma}=\beta_{\text{jmn1}}\,,\qquad \beta_{\text{jmn1}}=\frac{\ell}{\rho}-b_s\rho (M_0\rho)^\alpha \,,
\end{align}
where $g_{tt}=-(1-M_0)(M_0\rho)^{M_0/(1-M_0)}$ and $g_{rr}=1/(1-M_0)$ are given in~\eqref{JMN-1}. Note that at the matching surface $r=r_b$, we have $M_0\rho =1\to\rho =1/M_0$.\\

The motion of charged particles in the $\theta=\pi/2$ plane is said to be of repulsive (resp. attractive) nature if the magnetic force on the particle is repulsive (resp. attractive). As seen from above the $\theta=\pi/2$ plane, the motion of a positive charge (resp. negative charge) is repulsive, if it rotates counter-clockwise (resp. clockwise); the motion is attractive if one reverses the direction of rotation. Since the sign of $d\varphi/d\sigma$ may not remain constant during the motion, the character of the latter changes along the particle's path. We will not restrict ourselves to circular motion and the momentum 4-vector takes the form
\begin{align}
\label{F5}&	p^\mu = m\gamma (e_t{}^\mu +ue_r{}^\mu+ve_\varphi{}^\mu),\\
& \gamma = \frac{\mathcal{E}}{\sqrt{-g_{tt}}},\quad u^2=\frac{\gamma^2 -1-\beta^2}{\gamma^2},\quad v=\frac{\beta}{\gamma},\nonumber
\end{align}
where $\beta$ is either $\beta_{\text{Sch}}$ or $\beta_{\text{jmn1}}$ and $g_{tt}$ is either the Schwarzschild metric or JMN-1 metric.\\

Next, we consider the motion of two \emph{test} particles of same mass and opposite charges.  We assume that at some given moment the two particles have the same value of the $\rho$ coordinate, the same energy $\mathcal{E}$, and equal but opposite values of $\ell$. Since they already have opposite values of $b_s$~\eqref{F0}, this results in opposite values of $\beta_{\text{Sch}}$~\eqref{F2} [respectively of $\beta_{\text{jmn1}}$~\eqref{F4}] and of $d\varphi/d\sigma$, and implies a fully symmetric motion of the particles; motion of same nature (for both particles) keeping the signs of $d\varphi/d\sigma$ and $v$~\eqref{F5} opposite, but same values, during the subsequent motion of the particles and keeping as well the same values and signs of the radial velocity $u$~\eqref{F5}. Thus, the total linear momentum of the system of the two particles is just $P^\mu=2m\gamma (e_t{}^\mu +ue_r{}^\mu)$ yielding $-P^\mu P_\mu=4m^2(1+\beta^2)$, where $\beta$ is either $\beta_{\text{Sch}}$ or $\beta_{\text{jmn1}}$. The center-of-mass energy per particle mass is
\begin{equation}\label{F6}
\mathcal{E}_{CMs}=2\sqrt{1+\beta_{\text{Sch}}^2}\,,
\end{equation}
for a Schwarzschild black hole immersed in a magnetic field, and
\begin{equation}\label{F7}
\mathcal{E}_{CMj}=2\sqrt{1+\beta_{\text{jmn1}}^2}\,,
\end{equation}
for the JMN-1 naked singularity immersed in a magnetic field. In both expressions $\rho$ is constrained by requiring that the right-hand side of~\eqref{F1} and of~\eqref{F3} be positive. As noted earlier, at the matching surface $r=r_b$, we have $M_0\rho =1$, so that the functions $\beta_{\text{Sch}}$ and $\beta_{\text{jmn1}}$ and, consequently, the functions $\mathcal{E}_{CMs}$ and $\mathcal{E}_{CMj}$ are continuous at $r=r_b$ (corresponding to $\rho=1/M_0$), but their derivatives with respect to $\rho$, that is their rates of change, are discontinuous there. \\

The four panels of Fig.~\ref{figCM1} and Fig.~\ref{figCM2} depict $\mathcal{E}_{CMs}$ and $\mathcal{E}_{CMj}$ for different values of the parameters taking $M_0=0.5$ in Fig.~\ref{figCM1} and $M_0 =0.3$ (resp. $M_0 =0.8$) in the left (resp. right) panel of Fig.~\ref{figCM2}. For a Schwarzschild black hole immersed in a magnetic field, a blue (resp. red) plot corresponds to repulsive (resp. attractive) motion of the two particles. For the JMN-1 naked singularity, a purple (resp. black) plot corresponds to repulsive (resp. attractive) motion of the two particles. A dashed plot, which has the same shape for all values of the parameters, corresponds to a JMN-1 singularity (internal metric) and a Schwarzschild black hole (external metric) with no external magnetic field ($b_s=B=0$); this plot diverges as $\rho\to 0$ and extends from $\rho=0$ to $\infty$.\\

In Fig.~\ref{figCM1} the matching surface $r=r_b$ corresponds to $\rho =2$.In the left panel of Fig.~\ref{figCM1}, due to the values of the parameters ($\mathcal{E},\,\ell,\,b_s$), the particles are confined in the region $1.262\leq \rho\leq 6.966$. In the right panel of Fig.~\ref{figCM1} the particles are confined in the region $0< \rho\leq 1.963<2$ inside the naked singularity and the purple plot diverges as $\rho\to 0$.\\ 

In the left panel of Fig.~\ref{figCM2} the matching surface $r=r_b$ corresponds to $\rho =10/3$ and in the right panel the matching surface is at $\rho =10/8$. In Fig.~\ref{figCM2} the motion is constrained by $0.944\leq  \rho\leq 19.47$ (left panel) and $0< \rho\leq 19.47$ (right panel). In the right panel, the purple plot diverges as $\rho\to 0$.\\

There is a mild discontinuity in the tagent line to the graph of the center-of-mass energy at $\rho=1/M_0$ (corresponding to the matching surface $r=r_b$). $\rho=1/M_0$ is the intersection point of the purple and blue curves in Figs.~\ref{figCM1} \& ~\ref{figCM2}. This is because the derivatives of $\mathcal{E}_{CMs}$ and $\mathcal{E}_{CMj}$ are discontinuous at $\rho=1/M_0$, as noted earlier. Note that for the attractive case (red or black plots) the center-of-mass energy is an increasing function of $\rho$ and decreasing for the repulsive motion (purple or black plots).

\section{Discussion and Conclusions}\label{Discussion}
The behavior of magnetic fields in ultra-dense compact objects is a topic of great interest as many sophisticated high energy phenomena can be defined for such scenario.  In view of this, we first present a mathematical formalism to define electromagnetic four potential, electromagnetic field tensor and magnetic field components for any general spherically symmetric spacetime.  We also investigate the behavior of magnetic field in the JMN-1 singularity, which forms as a gravitational collapse of inhomogeneous matter cloud with anisotropic pressure. In the equilibrium state, JMN-1 singularity consists of matter up to radius $R_b$ while the exterior geometry of the structure is represented by Schwarzschild spacetime.\\

We have determined the master equation~\eqref{L2} and master differential equation~\eqref{L3} that determine the spatial behavior of the magnetic field inside a matter distribution or vacuum region, of general spherically symmetric metric, which is immersed in a test magnetic field. Then, we specialized to the case of JMN-1 singularity in a uniform weak magnetic field and we determined the electromagnetic four potential vector using a method which is a generalization of the Wald method. This allows the description of magnetic fields in vacuum and non-vacuum spacetimes. We also studied collision of particles in the vicinity of the magnetized JMN-1 singularity and analyze the behavior of center0if-mass energy of the colliding particles. Below we summarize the conclusions of this study:

\begin{itemize}

\item For the interior spacetime described by \eqref{JMN-1}, the non-vanishing component of the electromagnetic four potential $A^\mu$ depends on the metric parameter $M_0$ (or on the newly defined parameter $\alpha$) and the boundary radius or the matching radius $r_b$. Here, $M_0$ can have any value in the range $0<M_0<\frac{4}{5}$ ( $0<\alpha<0.701562$). The behavior of electromagnetic four potential is continuous at any radial distance $r$, including at the boundary radius $r_b$. For $r>r_b$, the expression of electromagnetic four potential takes the usual form of $A^\mu$ of the Schwarzschild spacetime. 

\item The orthogonal components of the magnetic field $B_{\hat{r}}$ and $B_{\hat{\theta}}$ also depend on the spacetime metric parameter $M_0$ and the boundary radius $r_b$. The normal component of the magnetic field $B_{\hat{r}}$ is continuous across the spacetime time geometry. However, for the physically realistic scenario, we find that the tangential component of the magnetic field $B_{\hat{\theta}}$ is discontinuous at $r=r_b$. These orthogonal magnetic field components are finite at $r=0$.

\item From magnetostatics, one can understand that, the magnetic field can be discontinuous at any given surface if the media present on both sides of boundary possess different properties. In the case of JMN-1, the discontinuity is observed at the boundary radius or matching radius $r_b$, where interior spacetime and exterior spacetime are matched. The interior spacetime consists of anisotropic collapsing matter, while there is vacuum in the exterior Schwarzschild spacetime. Since the magnetic permeability of the matter is different from that of the vacuum, the tangential component of the magnetic field is discontinuous at the boundary separating these two media. 


\item We also determine the covariant expression of the surface current on the surface of the matter of interior spacetime. This indicates that the surface current density depends on the magnetic permeabilities of the interior matter region and exterior vacuum region. The concept of surface currents in the context of astrophysics has been investigated in structures such as neutron stars \cite{Gralla2014,Gurevich, deSouza2018}, pulsar \cite{Zhang2017} etc. In many papers such as \cite{Gurevich}, many sophisticated processes are proposed to explain high energy phenomena in the context of neutron stars. Other than this, discontinuity of magnetic fields and existence of surface currents are very important aspects of Magnetohydrodynamics. The appearance of these features in compact objects such as horizonless singularities or interior solutions, which could be alternative to black holes, leads to multiple interesting possibilities that can be explored in future.

\item The study of collision of particles in the vicinity of the magnetized JMN-1 singularity shows that the center-of-mass energy of two oppositely charged particles assumes a continuous value across the separating surface (where the magnetic field is discontinuous), and that its radial rate of change observes a mild finite discontinuity. This could be a possible way to distinguish structures with discontinuous matter fields such as JMN-1 from spacetimes with continuous, or no, matter fields.

\end{itemize}

The general formalism presented in this paper to define magnetic field quantities for any given general spherically symmetric spacetimes will be important also for future studied. It will be useful to describe the behavior of magnetic fields in structures with continuous or discontinuous matter fluids.\\

The behavior of magnetic fields in the JMN-1 singularity is very interesting. In the existing studies of magnetic fields in non-vacuum compact objects, matter fluids exist in the whole universe and are continuous everywhere in the spacetime. While for JMN-1 singularity, the cloud in the equilibrium state consists inhomogenenous distribution of matter from the centre of the star to some finite radius. This region of spacetime represented by \eqref{JMN-1}, is matched with an exterior Schwarzschild spacetime. For this reason, JMN-1 singularity can possibly be considered as a well motivated physically realistic model of a galaxy. It is well known that inclusion of magnetic fields in many phenomena is quite important in astrophysical scenarios. Thus this study of magnetic fields in JMN-1 spacetime would be important in understanding many different phenomena in which dynamics of charged particles or plasma are important aspects. Other than this, this work will serve as a basis to investigate various phenomena such as prospects of Innermost Stable Circular Orbits of charged particles, motion of plasma, effects of magnetic fields on the shadow in the context of JMN-1 singularity, and polarized shadow images etc. \\

Apart from this, the feature of discontinuity in the magnetic fields in astrophysical structures which consists of matter fields separated by a boundary will be useful for future works. One can expect such discontinuity in these type of geometries. The investigation of phenomena associated with magnetic fields in different models can potentially serve as a good method to distinguish between the two types of compact objects, i.e. structures represented by continuous matter fields with the structure representing a fine separation between the matter fluids. The expected similar behavior of magnetic fields having discontinuity in different interior solutions opens multiple fascinating opportunities to explore different high energy emitting phenomena.\\

Only the ordinary (non-exotic) macroscopic properties of the anisotropic matter (energy density and tangential pressure), of which the JMN-1 singularity is made, have been used to derive the metric~\eqref{JMN-1} in \cite{Joshi:2011zm}. In the present manuscript we have used another ordinary macroscopic property (the permeability) to write down an expression for the discontinuity of the tangential component of the magnetic field across the boundary at $r=r_b$. We have not advanced any hypothesis about the microscopic structure of the JMN-1 singularity; however, since the energy density and tangential pressure are both positive [see the unnumbered equation following Eq.~\eqref{JMN-1}], we assume that the JMN-1 singularity is made of ordinary matter.

\section*{Appendix: Checking the validity of the interior solution for the magnetic field}
\appendix
\def\theequation{A.\arabic{equation}}
\setcounter{equation}{0}
With the only non-vanishing components of the electromagnetic field tensor $F_{\mu\nu}$ given in~\eqref{Frphi final} and~\eqref{Ftheta phi final} for the interior spacetime $(r\leq r_b)$, we check validity of the field equations $F_{[\mu \nu;\gamma]}=F_{[\mu \nu,\gamma]}=0$, which reduce to $F_{r\varphi,\theta}+F_{\varphi\theta,r}=0$. We have
\begin{align*}
&F_{r\varphi,\theta}=
(2 + \alpha) B r \left(\frac{r}{r_b}\right)^\alpha  \sin{\theta}\cos{\theta},\\
&F_{\varphi\theta,r}=-(2 + \alpha) B r \left(\frac{r}{r_b}\right)^\alpha  \sin{\theta}\cos{\theta}.
\end{align*}
Thus, $F_{r\varphi,\theta}+F_{\varphi\theta,r}=0$ is satisfied for all $\alpha$.

Next, we check the validity of $(\sqrt{|g|}F^{\mu\nu}){}_{,\nu}=0$. The only non-vanishing components of $F^{\mu\nu}$ are
\begin{align*}
&F^{r\varphi}=\frac{2 + \alpha}{2}~\frac{B(1-M_0)}{r}~\left(\frac{r}{r_b}\right)^\alpha ,\\
&F^{\theta\varphi}=\frac{B\cot\theta}{r^2}~\left(\frac{r}{r_b}\right)^\alpha .
\end{align*}
The equations $(\sqrt{|g|}F^{\mu\nu}){}_{,\nu}=0$ reduce to $(\sqrt{|g|}F^{\varphi r}){}_{,r}+(\sqrt{|g|}F^{\varphi \theta}){}_{,\theta}=0$, with $\sqrt{|g|}$ given by
\begin{equation*}
\sqrt{|g|}=r^2\sqrt{\left(\frac{r}{r_b}\right)^{\frac{M_0}{1-M_0}}}~~\sin\theta ,	
\end{equation*}
we obtain
\begin{multline}\label{A1}
(\sqrt{|g|}F^{\varphi r}){}_{,r}+(\sqrt{|g|}F^{\varphi \theta}){}_{,\theta}=\\
-\frac{B}{4}\sqrt{\left(\frac{r}{r_b}\right)^{\frac{M_0}{1-M_0}}}~\left(\frac{r}{r_b}\right)^\alpha [2(1-M_0)\alpha^2 + (6-5M_0)\alpha - 2M_0].
\end{multline}
The expression inside the square bracket in~\eqref{A1}, $2(1-M_0)\alpha^2 + (6-5M_0)\alpha - 2M_0$, is 0 for $\alpha$ given by~\eqref{alpha}.\\
\vskip7pt

\section*{Acknowledgments}

KA would like to thank Martin Kolo\v{s} and Arman Tursunov for a discussion which led to the idea of this work.


\begin{thebibliography}{99}

\bibitem{Penrose:1971uk}
R.~Penrose and R.~M.~Floyd,
Nature \textbf{229} (1971), 177-179

\bibitem{Tursunov:2019oiq}
A.~Tursunov and N.~Dadhich,
Universe \textbf{5} (2019) no.5, 125

\bibitem{Blandford:1977ds}
R.~D.~Blandford and R.~L.~Znajek,
Mon. Not. Roy. Astron. Soc. \textbf{179} (1977), 433-456


\bibitem{Tursunov:2018erf}
A.~Tursunov, M.~Kolo\v{s}, Z.~Stuchl\'\i{}k and D.~V.~Gal'tsov,
Astrophys. J. \textbf{861} (2018) no.1, 2

\bibitem{Ferriere_2015}
K.~Ferri\`{e}re,
J. Phys.: Conf. Ser. \textbf{577}, 012008 (2015)

\bibitem{Stuchlik2000}Z.~Stuchl\'\i{}k, M.~Kolo\v{s}, J.~Kov\'{a}\v{r}, P.~Slan\'{y} and A.~Tursunov, Universe \textbf{2000}, 6(2), 26 

\bibitem{ansatz1}R.~Wald, Phys. Rev. D {\bf 10}, 1680 (1974)	

\bibitem{Joshi:1993zg}
P.~S.~Joshi and I.~H.~Dwivedi,
Phys. Rev. D \textbf{47} (1993), 5357-5369

\bibitem{Dwivedi:1994qs}
I.~H.~Dwivedi and P.~S.~Joshi,
Commun. Math. Phys. \textbf{166} (1994), 117-128

\bibitem{Singh:1994tb}
T.~P.~Singh and P.~S.~Joshi,
Class. Quant. Grav. \textbf{13} (1996), 559-572

\bibitem{Saurabh:2023otl}
Saurabh, P.~Bambhaniya and P.~S.~Joshi,
[arXiv:2308.14519 [astro-ph.HE]].

\bibitem{Vagnozzi:2022moj}
S.~Vagnozzi, R.~Roy, Y.~D.~Tsai, L.~Visinelli, M.~Afrin, A.~Allahyari, P.~Bambhaniya, D.~Dey, S.~G.~Ghosh and P.~S.~Joshi, \textit{et al.}
Class. Quant. Grav. \textbf{40} (2023) no.16, 165007

\bibitem{EventHorizonTelescope:2022xqj}
K.~Akiyama \textit{et al.} [Event Horizon Telescope],
Astrophys. J. Lett. \textbf{930} (2022) no.2, L17

\bibitem{Bambhaniya:2019pbr}
P.~Bambhaniya, A.~B.~Joshi, D.~Dey and P.~S.~Joshi,
Phys. Rev. D \textbf{100} (2019) no.12, 124020

\bibitem{Joshi:2020tlq}
A.~B.~Joshi, D.~Dey, P.~S.~Joshi and P.~Bambhaniya,
Phys. Rev. D \textbf{102} (2020) no.2, 024022


\bibitem{doran2006}
R.~Doran, F.~S.~N.~Lobo and P.~Crawford,
Found. Phys. \textbf{38} (2008), 160-187

\bibitem{Mazur2004}
P.~O.~Mazur and E.~Mottola,
Proc. Nat. Acad. Sci. \textbf{101} (2004), 9545-9550


\bibitem{Chapline2004}
G.~Chapline,
eConf \textbf{C041213} (2004), 0205


\bibitem{Joshi:2011zm}P.S.~Joshi, D.~Malafarina and R.~Narayan, Class. Quantum. Grav. {\bf 28}, 235018 (2011)


\bibitem{ansatz2}M.~Azreg-A\"{\i}nou, Eur. Phys. J. C {\bf 76}, 414 (2016)

\bibitem{Oppenheimer:1939ue}
J.~R.~Oppenheimer and H.~Snyder,
Phys. Rev. \textbf{56}, 455-459 (1939)

\bibitem{mosani3}
K.~Mosani, D.~Dey and P.~S.~Joshi,
Monthly Notices of the Royal Astronomical Society, \textbf{ 504}, 4, 4743–4750 (2021)

\bibitem{Mosani:2022nsp}
K.~Mosani,
[arXiv:2211.06604 [gr-qc]]

\bibitem{Shaikh2018lcc}
R.~Shaikh, P.~Kocherlakota, R.~Narayan and P.~S.~Joshi,
Mon. Not. Roy. Astron. Soc. \textbf{482} (2019) no.1, 52-64

\bibitem{Shaikhhbm}
R.~Shaikh and P.~S.~Joshi,
JCAP \textbf{10} (2019), 064

\bibitem{PRD}
A.~M.~Al Zahrani, V.~P~Frolov and A.~A.~Shoom,
Phys. Rev. D \textbf{87}, 084043 (2013)

\bibitem{CQG}
M.~Kolo\v{s}, Z~Stuchl\'{\i}k and A.~Tursunov,
Class. Quantum. Grav. {\bf 32}, 165009 (2015)

\bibitem{handbook}A.D.~Polyanin and V.F.~Zaitsev, \textit{Handbook of Ordinary Differential Equations} (CRC Press, Taylor \& Francis Group, Boca Raton, 2018)

\bibitem{Hayward}
Z.~Yang and Y.~K.~Lim, Phys. Rev. D \textbf{105}, no.12, 124045 (2022)




\bibitem{Gammametric}
C.~A.~Benavides-Gallego, A.~Abdujabbarov, D.~Malafarina and C.~Bambi, Phys. Rev. D \textbf{101}, no.12, 124024 (2020)

\bibitem{ErnstSch}
D.~Li and X.~Wu, Eur. Phys. J. Plus \textbf{134}, no.3, 96 (2019)

\bibitem{sun}
Y.-J.~Moon \textit{et. al.}, Solar Physics \textbf{184}, 323 (1999)



\bibitem{Gralla2014}
S.~E.~Gralla and T.~Jacobson,
Mon. Not. Roy. Astron. Soc. \textbf{445} (2014) no.3, 2500-2534



\bibitem{Gurevich}
A.~V.~Gurevich and N.~Ya.~Istomin,
Mon. Not. Roy. Astron. Soc. \textbf{377}, 4, 1663-1667 (2007) 



\bibitem{deSouza2018}
G.~H.~de Souza and C.~Chirenti,
Phys. Rev. D \textbf{100} (2019) no.4, 043017



\bibitem{Frolov}
V.~P.~Frolov,
Phys. Rev. D \textbf{85}, 024020 (2012)


\bibitem{Zhang2017}
F.~Zhang,
Astron. Astrophys. \textbf{598} (2017), A88


\end{thebibliography}
\end{document}